\documentclass[pra,superscriptaddress,twocolumn,english, longbibliography, showpacs=false, nofootinbib]{revtex4-2}

\usepackage[T1]{fontenc}
\usepackage[utf8]{inputenc}
\usepackage{xcolor}
\usepackage{babel}
\usepackage{amsmath}
\usepackage{amssymb}
\usepackage{subfigure}
\usepackage{graphicx}
\usepackage{physics} 
\usepackage{dsfont}
\usepackage{hyperref}
\usepackage{amsthm}
\usepackage{mathtools}
\usepackage[left=23mm,right=13mm,top=35mm,columnsep=15pt]{geometry} 
\usepackage{adjustbox}
\usepackage{placeins}
\usepackage{csquotes}
\usepackage{mathrsfs}
\usepackage{enumerate}
\usepackage{verbatim}

\begin{document}

\title{Entropy production due to spacetime fluctuations}

\author{Thiago H. Moreira\href{https://orcid.org/0000-0001-7093-0287}{\includegraphics[scale=0.05]{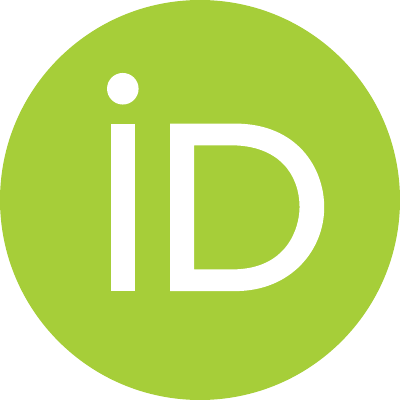}}}
\email{thiagohenriquemoreira@ufg.br}
\affiliation{QPequi Group, Institute of Physics, Federal University of Goi\'as, Goi\^ania, Goi\'as, 74.690-900, Brazil}

\author{Lucas C. Céleri\href{https://orcid.org/0000-0001-5120-8176}{\includegraphics[scale=0.05]{orcidid.pdf}}}
\email{lucas@qpequi.com}
\affiliation{QPequi Group, Institute of Physics, Federal University of Goi\'as, Goi\^ania, Goi\'as, 74.690-900, Brazil}

\begin{abstract}
Understanding the quantum nature of the gravitational field is undoubtedly one of the greatest challenges in theoretical physics. Despite significant progress, a complete and consistent theory remains elusive. However, in the weak field approximation ---where curvature effects are small--- we can explore some expected properties of such a theory. Particularly relevant to this study is the quantum nature of gravitational waves, which are represented as small perturbations in flat spacetime. In this framework, a quantum description of these perturbations, as a quantum field, is feasible, leading to the emergence of the graviton. Here we consider a non-relativistic quantum system interacting with such a field. We employ the consistent histories approach to quantum mechanics, which allows us to frame classical questions in a quantum context, to define a fluctuation relation for this system. As a result, thermodynamic entropy must be produced in the system due to its unavoidable interaction with the quantum fluctuations of spacetime. 
\end{abstract}

\maketitle


\section{Introduction}

Fluctuation theorems are strong results in nonequilibrium statistical mechanics, with applications spanning a wide variety of fields. These range from human learning~\cite{Hack2023} and cosmology~\cite{Chevalier2007} to biology~\cite{Liphardt2002}, chemistry~\cite{Vlad1998}, and quantum field theory~\cite{Bartolotta2008}, to name just a few. Essentially, these theorems illustrate a strong connection between equilibrium quantities such as Helmholtz free energy and non-equilibrium ones, like energy and work. While various versions of these theorems exist, the most renowned are those formulated by Jarzynski~\cite{Jarzynski1997} and Crooks~\cite{Crooks1999}. For a historical overview of this topic, see Ref.~\cite{Darrigol2023}, and for technical details on the general theorems, consult Refs.~\cite{Jarzynski2008,Jarzynski2011,Seifert2012,Esposito2009,Campisi2011}.

In the development of quantum field theory, one key implication is the inherent fluctuation of the quantum field energy. These fluctuations underpin various physical phenomena, such as the Casimir effect~\cite{Casimir1948} and the Lamb shift~\cite{Welton1948}. When a system interacts with such fields and attempts to change these fluctuations, it induces frictional forces, leading to dissipation and entropy production~\cite{Kardar1999,Oliveira2024}.

In this context, a particularly intriguing aspect is the quantum nature of the gravitational field. Although a complete theory of quantum gravity remains elusive, it is possible to construct a consistent quantum field theory in the linear regime of Einstein’s theory, where curvature is small. This approach, known as the weak field approximation, introduces gravitons as the particles associated with the gravitational field~\cite{Weiberg1972}. The primary aim of this work is to examine entropy production in a system interacting with a weak quantum gravitational field. Specifically, by modeling the system as an open quantum system interacting with a bath of gravitons, we derive a fluctuation theorem for the system, the first moment of which is the entropy production. It is interesting to observe here that, contrary to other fields where energy fluctuates on a well defined spacetime, the quantum nature of the gravitational waves introduces fluctuations of the spacetime itself. We note that the effects of the curvature on the entropy production is not being taking into account here. This classical contribution to entropy was recently considered in Refs.~\cite{Basso2024,Basso2023}.

One major conceptual challenge in formulating a quantum fluctuation theorem is defining work. In classical systems, work is a well-defined quantity dependent on the trajectory and does not fluctuate, unless the system is immerse in a fluctuating environment, like a heat bath. In such cases, stochastic thermodynamics is required to describe the system and physical quantities emerge as averages over the fluctuations. In quantum systems, fluctuations are unavoidable, even in isolated systems. Moreover, the lack of the notion of a trajectory for a quantum system makes unambiguous the extension of the notion of thermodynamic quantities, like work.

For closed quantum systems, the definition of work using a two-time measurement scheme~\cite{Talkner2007} is well-established. However, defining work for open quantum systems remains challenging. In this paper, we adopt the approach proposed in Ref.~\cite{Hu2012}, which defines work in a way analogous to classical systems by interpreting trajectories within quantum mechanics. To achieve this, we employ the decoherent histories formalism, primarily developed by Griffiths, Omnès, Gell-Mann, and Hartle~\cite{Griffiths1984,Omnes1990,Omnes1992,Gell-Mann-Hartle1990}, with additional insights from Ref.~\cite{Dowker1992}. This formalism was designed to describe quantum systems without relying on external observers. It employs Schrödinger's equation to help assigning probabilities to different histories, allowing us to consistently address classical questions within a quantum framework. One such question is the definition of work, which we approach by considering individual quantum histories, keeping in mind the classical thermodynamic idea of a trajectory dependent quantity.

The remaining of the paper is structured as follows. First, we define our system of interest in the next section, Sec.~\ref{sec:system}, where the notation is also established. The decoherent histories approach to quantum mechanics is reviewed in Sec.~\ref{sec:decoherent}. The main result, the fluctuation theorem, is presented in Sec.~\ref{sec:fluctuation}, where entropy production is directly related to quantum fluctuations of the spacetime. Finally, Sec.~\ref{sec:discussion} presents a physical discussion and our concluding remarks. We employ natural units in which $\hbar=c=k_B=G=1$ and the metric signature $(-,+,+,+)$.

\section{Localized quantum system in linearized quantum gravity}
\label{sec:system}

Our system of interest consists of two massive particles interacting with a gravitational wave within a flat spacetime background. Upon quantization, the fluctuations of the gravitational waves introduce a stochastic dynamic to the particles. In this section, we provide a brief overview of the approach presented in Refs.~\cite{Parikh_2021,Parikh2021,Cho2022}, which derives a Langevin-like equation governing the geodesic separation between the two masses in the presence of the weak quantum gravitational field. For additional details, the reader is also referred to Ref.~\cite{Moreira2023}.

Let us start by writing the classical action for a weak gravitational field interacting with a pair of massive particles. The total action is given by $S = S_{\rm m} + S_{\rm EH}$, where $S_{\rm m}$ represents the action of the particles and their interaction with the gravitational field, and $S_{\rm EH}$ is the Einstein–Hilbert action that describes the gravitational field itself. Denoting the coordinates of the particles as $\zeta^\mu$ and $\xi^\mu$, the action for the particles is expressed as
\begin{equation}
\begin{split}
    S_{\rm m}&=-M\int \dd t\,\sqrt{-g_{\mu\nu}\dot{\zeta}^\mu\dot{\zeta}^\nu} \\
    &-m\int \dd t\,\sqrt{-g_{\mu\nu}\dot{\xi}^\mu\dot{\xi}^\nu}-V(\xi).
\end{split}
\end{equation}
We consider that the first particle ($M$) is free falling, while the second one ($m$) suffers the action of a potential $V$. Moreover, we take $M\gg m$ such that the heavier particle is on shell with worldline $\zeta_0^\mu(t)$. In this way, its dynamics becomes negligible. Furthermore, we take this particle to be at rest at the origin of our coordinate system, $\zeta_0^\mu(t)=t\delta_0^\mu$. Therefore, we can interpret the coordinate time $t$ as its proper time. Since our interest lies on the lighter particle, from here on we simply refer to it as the system.

Under these assumptions it becomes appropriate to think of $(t,\xi^i)$ as the Fermi normal coordinates defined with respect to the worldline of the heavier particle \cite{Manasse1963}. In these coordinates, we find \cite{Moreira2023}
\begin{equation}
    \sqrt{-g_{\mu\nu}(\xi)\dot{\xi}^\mu\dot{\xi}^\nu}\simeq1-\frac{1}{2}\delta_{ij}\dot{\xi}^i\dot{\xi}^j+\frac{1}{2}R_{i0j0}(t,0)\xi^i\xi^j
\end{equation}
in the nonrelativistic limit. Here, $R_{\mu\nu\rho\sigma}$ is the Riemann curvature tensor.

Now we write the spacetime metric as $g_{\mu\nu}=\eta_{\mu\nu}+h_{\mu\nu}$, with $\abs{h_{\mu\nu}}\ll 1$, where $\eta_{\mu\nu}=\rm{diag}(-1,1,1,1)$ is the flat Minkowski metric while $h_{\mu\nu}$ is a small perturbation. This is the weak field regime for the gravitational field, where the spacetime curvature is small compared with the relevant scales of the problem. Under this condition, the Einstein field equation reduces to a linear wave equation for $h_{\mu\nu}$. By taking advantage of the gauge invariance of linearized gravity, we choose to work in the transverse-traceless gauge such that the metric perturbation obeys $\Bar{h}_{0\nu}=0$, $\eta^{\mu\nu}\Bar{h}_{\mu\nu}=0$ and $\partial^\mu\Bar{h}_{\mu\nu}=0$, with $\Bar{h}_{\mu\nu} = h_{\mu\nu} - h\,\eta_{\mu\nu}/2$ and $h = h^{\mu}_{\mu}$. In this gauge, the non-vanishing components of the Riemann curvature tensor are given by $R_{0i0j}(t,0)=-\ddot{\Bar{h}}_{ij}(t,0)/2$, thus implying that the matter action takes the form
\begin{equation}
    S_{\rm m}=\int \dd t\,\qty[\frac{1}{2}m\,\delta_{ij}\dot{\xi}^i\dot{\xi}^j-V(\xi)+\frac{1}{4}m\,\ddot{\Bar{h}}_{ij}(t,0)\xi^i\xi^j],
\end{equation}
where we have dropped non-dynamical terms.

It can be shown that the coupling with the weak gravity field leads to the total action given by \cite{Moreira2023}
\begin{align} \label{Total-action}
    S&=\frac{\pi^{2}}{8}\int \dd t\int \dd^3k\,\sum_s\qty[\dot{q}^2_s(t,\vb{k})-\vb{k}^2q_s^2(t,\vb{k})] \nonumber \\
    &+\int \dd t\,\qty[\frac{1}{2}m\,\delta_{ij}\dot{\xi}^i\dot{\xi}^j-V(\xi)] \nonumber \\
    &+\frac{m}{4}\int \dd t\,\int \dd^3k\,\sum_sq_s(t,\vb{k})\,X^s(t,\vb{k}),
\end{align}
where we have defined
\begin{equation}
    X^s(t,\vb{k})\equiv\dv[2]{t}\qty[\epsilon_{ij}^s(\vb{k})\xi^i\xi^j].
\end{equation}
The gravitational field is expressed in terms of its two physical degrees of freedom, corresponding to the two polarization components by means of a Fourier transform
\begin{equation}
    \Bar{h}_{ij}(t,\vb{x})=\int \dd^3k\,\sum_s\epsilon_{ij}^{s}(\vb{k})q_s(t,\vb{k})e^{i\vb{k}\cdot\vb{x}},
\end{equation}
where $q_s(t,\vb{k})$ denotes the polarization-dependent field amplitudes in Fourier space and $\epsilon_{ij}^{s}$ denotes the polarization tensor that satisfies the normalization $\epsilon_{ij}^s(\vb{k})\epsilon^{ij}_{s'}(\vb{k})=2\delta^s_{s'}$, the transversality $k^i\epsilon_{ij}^{s}(\vb{k})=0$ and the traceless $\delta^{ij}\epsilon_{ij}^{s}(\vb{k})=0$ conditions. The coupling of the system with the field amplitudes is encoded by the variable $X^s(t,\vb{k})$.

Now, we assume that the initial state of the entire system is a product state, implying that initially there are no correlations between any of the components of the system ---the two particles and the field. Using the Feynman-Vernon influence functional approach to open quantum systems~\cite{Feynman1963,Feynman2010}, we can express all (weak) quantum gravitational influences on the system through the influence functional
\begin{align} \label{Influence-functional}
    &e^{iS_{\rm IF}[x,x']} = \int\mathcal{D}\mathcal{N}P[\mathcal{N}]e^{i\int \dd t\mathcal{N}^{ij}(t)\qty[x_{ij}(t)-x_{ij}'(t)]} \nonumber \\
    &\times e^{\frac{i}{2}\int \dd t\dd t'\,\qty[x_{ij}(t)-x_{ij}'(t)]D^{ijkl}(t,t')\qty[x_{kl}(t')+x_{kl}'(t')]},
\end{align}
where $x_{ij}(t)=\xi_i(t)\xi_j(t)$. Here, $\mathcal{N}^{ij}(t)$ is a stochastic variable with Gaussian probability density $P[\mathcal{N}]$ such that $\expval{\mathcal{N}^{ij}(t)}=0$ and $\expval{\mathcal{N}^{ij}(t)\mathcal{N}^{kl}(t')}=N^{ijkl}(t,t')$.
The quantities $D^{ijkl}(t,t')$ and $N^{ijkl}(t,t')$ are called the dissipation and noise kernels, respectively.

In summary, the system is found to be described by a stochastic effective action given by~\cite{Cho2022}
\begin{equation}
    S_{\rm SEA}=S_{\rm p}[\xi]-S_{\rm p}[\xi']+S_{\rm IF},
\end{equation}
with $S_{\rm p}[\xi]=\int \dd t\,\qty[\frac{1}{2}m\,\delta_{ij}\dot{\xi}^i\dot{\xi}^j-V(\xi)]$ and $S_{\rm IF}$ being the influence action defined in Eq.~\eqref{Influence-functional}. The equation of motion is obtained by setting $\delta S_{\rm SEA}=0$~\cite{Cho2022}, yielding
\begin{equation}
\begin{split}
    m\Ddot{\xi}_i(t)-2\int \dd t'\,D_{ijkl}(t,t')\xi^{j}(t)\xi^k(t')\xi^l(t') \\
    =-\pdv{V}{\xi^i}+2\mathcal{N}_{ij}(t)\xi^j(t).
\end{split}
\end{equation}
Note that the second term on the left-hand side introduces a non-linearity, making this integral differential equation very difficult to solve. It is for this reason that we choose to work on a perturbative regime, thus dropping the term involving the dissipation kernel as it was done in Ref.~\cite{Moreira2023}. In this regime, we obtain
\begin{equation} 
    m\Ddot{\xi}_i(t)=f_i(t)+2\mathcal{N}_{ij}(t)\xi_0^j(t),
\end{equation}
where $f_i$ is defined such that $V(\xi)=-f_i(t)\xi^i(t)$ and $\xi_0^i$ is the solution to $m\Ddot{\xi}_0^i=0$.

The above equation is a Langevin-like equation involving the stochastic force term $2\mathcal{N}_{ij}(t)\xi_0^j(t)$. The well-known solution for the initial conditions $\xi^i(0)=\dot{\xi}^i(0)=0$ is given by
\begin{equation} \label{Langevin-like-solution}
\begin{split}
    \xi_i(t)&=\frac{1}{m}\int_0^t\dd t'\,\qty(t-t')f_i(t') \\
    &+\frac{2}{m}\int_0^t \dd t'\,\qty(t-t')\mathcal{N}_{ij}(t')\xi_0^j(t').
\end{split}
\end{equation}
This can be viewed as the classical limit for the geodesic separation between the two masses, where the stochastic behavior arises from the quantum fluctuations of the gravitational field. The key observation here is that, under certain conditions, we can employ the solutions to the Langevin equation to define work for the quantum system similarly to how it is defined for classical systems. This approach relies significantly on the concept of trajectories in space. To explore how this is feasible, we will work within the consistent (or decoherent) histories approach to quantum mechanics, which tells us under what conditions the concept of trajectory for a quantum system becomes meaningful. Note that the system is still quantum in the sense that no classical limit is being considered in this approach. This is a crucial step that allows us to properly build a fluctuation theorem.

\section{Decoherent histories}
\label{sec:decoherent}

Shortly, a quantum mechanical history is a sequence of events at successive instants of time, being characterized by a sequence of successive projection operators. Let us begin by considering a closed quantum system initially described by a density matrix $\rho_0\equiv\rho(t_0)$ and we wish to consider histories consisting of $n$ projections at times $t_1<t_2<\dots<t_n$, with $t_{1}>t_0$. In order to establish whether or not we can assign probabilities to histories (and how we do this), we need to look at the so-called \emph{decoherence functional}. For a pair of histories, $[\alpha]$ and $[\beta]$, where $[\alpha]$ denotes the sequence of events $\lbrace \alpha_i\rbrace_{i=1}^{n}$ at times $\lbrace t_i\rbrace_{i=1}^{n}$, and analogously for $[\beta]$, the decoherence functional is defined as
\begin{equation}   
    \mathcal{D}\qty[\alpha,\beta]=\textrm{Tr}\qty[P_{\alpha_n}\dots P_{\alpha_1}\rho_0P_{\beta_1}\dots P_{\beta_n}],
\end{equation}
Here, $P_{\alpha_k}$ is some set of projection operators compatible with event $\alpha_k$ (at time $t_k$), $P_{\alpha_k}=U^\dagger(t_k,t_0)P_{\alpha_0}U(t_k,t_0)$, where $U(t_k,t_0)$ is the unitary time evolution operator for the system. Note that we can have different types of projections at different moments of time (although in this work we are mainly concerned with projections in position)~\cite{Dowker1992}. The decoherence functional can be written in a more compact form as~\cite{Calzetta2008}
\begin{equation} \label{Decoherence-functional}
    \mathcal{D}\qty[\alpha,\beta]=\textrm{Tr}\qty{\Tilde{T}\qty[\prod_{j=1}^nP_{\beta_j}]T\qty[\prod_{i=1}^nP_{\alpha_i}]\rho_0},
\end{equation}
where $T$ ($\Tilde{T}$) stands for time (anti-time) ordering. It is straightforward to show that the decoherence functional satisfies the following properties
\begin{enumerate}[i)]
    \item $\mathcal{D}\qty[\alpha,\beta]=\mathcal{D}^*\qty[\beta,\alpha]$,
    \item $\sum_{[\alpha],[\beta]}\mathcal{D}\qty[\alpha,\beta]=1$,
    \item $\mathcal{D}\qty[\alpha,\alpha]\geq0$,
    \item $\sum_{[\alpha]}\mathcal{D}\qty[\alpha,\alpha]=1$,
\end{enumerate}
where we are employing the shorthand notation $\sum_{[\alpha]}~\equiv ~\sum_{\alpha_1}\sum_{\alpha_2}\dots\sum_{\alpha_n}$.

The last two properties suggest that we identify the diagonal elements $\mathcal{D}[\alpha,\alpha]$ with the probability for the history $(\rho_0,t_0)\to\qty(\alpha_1,t_1)\to\dots\to\qty(\alpha_n,t_n)$, namely $p(\alpha)=\mathcal{D}[\alpha,\alpha]$~\cite{Dowker1992}. However, in order for such elements to represent a genuine probability, Kolmogorov's third axiom of probability theory ---the probability sum rule--- must be satisfied. It is not difficult to show that
\begin{align}
    p(\alpha\vee\beta)&=\mathcal{D}[\alpha,\alpha]+\mathcal{D}[\beta,\beta]+2\textrm{Re}\mathcal{D}[\alpha,\beta] \nonumber \\
    &=p(\alpha)+p(\beta)+2\textrm{Re}\mathcal{D}[\alpha,\beta].
\end{align}
Clearly, the sum rule is not obeyed in general due to the last term on the right-hand side. This term characterizes the quantum mechanical interference between distinct histories and it is in this sense that interference effects prevents us to give a meaningful notion of trajectory in quantum mechanics. However, this notion acquires meaning when there is strong decoherence, case in which $\mathcal{D}[\alpha,\beta]$ vanishes for distinct trajectories. One may then write the fundamental formula for the quantum mechanics of histories as
\begin{equation}
    \textrm{Re}\qty{\mathcal{D}[\alpha,\beta]}=p(\alpha)\delta_{\alpha_1,\beta_1}\dots\delta_{\alpha_n,\beta_n}.
\end{equation}
This equation provides the necessary and sufficient condition for assigning probabilities to individual histories, as well as specifies what those probabilities are. However, for most applications of the formalism, it is often sufficient to assert that a pair of mutually exclusive histories is consistent when $\textrm{Re}\qty{\mathcal{D}[\alpha,\beta]}<<\mathcal{D}[\alpha,\alpha],\,\mathcal{D}[\beta,\beta]$ for $[\alpha]\neq [\beta]$~\cite{Calzetta2008}.

From now on, let us consider projections in position basis, which is a kind of history implemented naturally in the path integral formalism. The projectors are represented by window functions $P_{\alpha_k}\to w_\alpha[x(t_k)]$, which are equal to one if the configuration at $t_k$ satisfies the requirement of history $[\alpha]$ and zero otherwise~\cite{Hu2012,Calzetta2008}. Equation~\eqref{Decoherence-functional} can then be written as
\begin{equation}
\begin{split}
    \mathcal{D}[\alpha,\beta]=&\int\mathcal{D}y\mathcal{D}y'\,e^{i\qty(S[y]-S[y'])}\rho\qty(y(0),y'(0),0) \\
    &\times\qty{\prod_iw_\alpha\qty[y(t_i)]}\qty{\prod_jw_\beta\qty[y'(t_j)]}.
\end{split}
\end{equation}
Here $S$ is the action for the total system described by some generic coordinate $y$.

Let us apply this to the problem at hand. By factoring the total action as shown in Eq.~\eqref{Total-action}, the integration over the gravitational variables leads to the influence functional given in Eq.~\eqref{Influence-functional}. We can then consider $[\alpha]$ and $[\beta]$ as histories represented by two non-overlapping paths $\xi^{(1)}(t)$ and $\xi^{(2)}(t)$, in which case the window functions would be simply delta functions at each time instant. For this scenario, the decoherence functional was analyzed in Ref.~\cite{Moreira2023}, considering four possible types of initial graviton states and estimating the decoherence time as a function of the gravitational energy scale (either a cutoff $\Lambda_{\rm g}\sim \xi^{-1}$ \cite{Kanno2021} or the temperature $\beta^{-1}$ of a thermal state of gravitons)\footnote{The analysis done in Ref.~\cite{Moreira2023} considered that the particle interacting with the gravitons was described by additional dynamical internal degrees of freedom that also coupled with the gravitational field. This coupling turned out to increase even more the decoherence rate. Although this is a feature to be taken advantage of, the internal degrees of freedom noise kernel also introduces a non-trivial modification to the equation of motion.}. Decoherence induced by gravitons was also analyzed in Ref. \cite{Kanno2021}, where it was established that decoherence occurs for particles with mass much greater than the Planck mass $M_{\rm pl}\sim10^{-8}$ kg. For such particles, this allows us to give meaning to the concept of trajectory for the quantum system since the interference terms may be neglected due to decoherence. Within this scenario we derive a quantum fluctuation theorem for our system. 

\section{Fluctuation theorem}
\label{sec:fluctuation}

In order to discuss the validity of the fluctuation theorem, we proceed as in Ref.~\cite{Hu2012}. By fluctuation theorem we mean Jarzynski-like relation, which is the integral form of Crook's fluctuation theorem~\cite{Jarzynski1997,Crooks1999,Jarzynski2007,Talkner2007,Jarzynski2008,Campisi2011}. For closed quantum systems, the derivation of the theorem relies on the hypothesis of initial thermal state of the entire system. Work is then defined as the difference in the energy of the system, measured before and after a given force protocol has acted upon it. Together with the sampling from an initial thermal state, the inherent quantum mechanical nature of the energy measurements yields a probabilistic nature to the work distribution. Under these conditions, the integrated fluctuation theorem states that~\cite{Jarzynski1997}
\begin{equation}
    \expval{e^{-\beta w}}=e^{-\beta\Delta F},
\end{equation}
where the average is performed using the work probability distribution, $\beta$ is the inverse initial temperature and $\Delta F$ is the free energy difference between final and initial equilibrium states.

However, our analysis of a quantum particle interacting with the gravitons was built under the assumption of initial product state, which is not a thermal state. To circumvent this issue, we take the initial time to be $t_0=-\infty$. At this time instant, we assume that the total state of the system is described by a tensor product, $\rho(-\infty)=\rho_{\rm sys}(-\infty)\otimes\rho_{\rm grav}(-\infty)$, where $\rho_{\rm grav}$ is a thermal state of gravitons. We then allow the system to evolve according to the total action with $f^i(t)=f^i(0)$ for $t<0$, so that we obtain a total thermal state at $t=0$. At $t=0$, the driving force starts changing according to some arbitrary protocol until a given time instant $t=\tau$. Assuming the decoherence conditions discussed in the previous section hold, the work performed on the system in the interval $\qty[0,\tau]$ is defined as
\begin{equation}
    W\equiv-\int_0^\tau \dd t\,\dot{f}_i(t)\xi^i(t),
\end{equation}
where $\xi^i(t)$ is the solution to the Langevin-like equation given in Eq.~\eqref{Langevin-like-solution}. Physically, this is the work that must be performed on the system in order to move it through the bath of gravitons. By using the explicit forms of the solutions we can write $W = W_f + W_n$, with
\begin{equation}
    W_f=-\frac{1}{m}\int_0^\tau \dd t \dd t'\,\dot{f}^i(t)g(t-t')f_i(t'),
    \label{eq:rev_work}
\end{equation}
denoting the work associated with the driven protocol, while 
\begin{equation}
    W_n = -\frac{2}{m}\int_0^\tau \dd t\dd t'\,\dot{f}^i(t)g(t-t')\mathcal{N}_{ij}(t')\xi^j_0(t'),
\end{equation}
contains the quantum fluctuations coming from the noise. In these equations, $g(t-t')=\qty(t-t')\theta(t-t')$, with $\theta(x)$ being the step function. 

While $W_f$ is entirely determined by the driven protocol, which is arbitrary, $W_n$ encompasses both the driven contribution, as it includes $f$, and the noise term resulting from the interaction with the gravitons. Since gravity cannot be shielded and its quantum nature inherently involves fluctuations, this latter contribution, unlike the one arising from the driven protocol, is unavoidable.

We note that the work performed on the system is linear in $\mathcal{N}_{ij}(t)$, which is a Gaussian random process. Consequently, $W$ itself must also be a Gaussian random variable. Its statistics are characterized by the first two moments $\expval{W}$ and $\sigma_W^2=\expval{W^2}-\expval{W}^2$. Therefore, we can write
\begin{equation}
    P(W)=\frac{1}{\sqrt{2\pi\sigma_W^2}}e^{-\qty(W-\expval{W})^2/2\sigma_W^2}.
\end{equation}
The brackets denote stochastic average with a Gaussian probability density.

Before we proceed, it is important to emphasize that the average with $P(W)$ is equivalent to the stochastic average with $P(\mathcal{N})$ due to the linear dependence of $W$ on $\mathcal{N}_{ij}(t)$. In conventional formulations of the classical fluctuation theorem, such stochastic forces are not typically considered, and the probabilistic nature of the work done on the system arises from the assumption of an initial thermal state. Each sampling from the initial state produces a trajectory, and the average is taken over an ensemble of such realizations. For the quantum fluctuation theorem, the inherent quantum uncertainty of the quantum state introduces an additional probabilistic aspect. However, we have opted to use an equivalent initial state preparation method based on a product initial state for the total system. This choice replaces the system's dependence on the initial state with the properties of noise statistics, thereby leaving only one probabilistic element instead of two. For further discussion on this topic, see Ref.~\cite{Hu2012}.

The moments of the work distribution can be directly computed as follows. First, since $\expval{\mathcal{N}^{ij}(t)}=0$, we have
\begin{equation} \label{First-moment-of-work}
    \expval{W}=-\frac{1}{m}\int_0^\tau dtdt'\,\dot{f}^i(t)g(t-t')f_i(t') = \expval{W_f}.
\end{equation}
Additionally, by using $\expval{\mathcal{N}^{ij}(t)\mathcal{N}^{kl}(t')}=N^{ijkl}(t,t')$, which is the noise kernel of gravitons, we also find
\begin{equation} \label{Second-moment-of-work}
    \sigma_W^2=\int_0^\tau dtdt'\,\dot{f}^i(t)\sigma_{ij}(t,t')\dot{f}^j(t'),
\end{equation}
where
\begin{eqnarray} \label{sigma_ij}
    \sigma_{ij}(t,t')&=&\frac{4}{m}\int_0^\tau dt_1dt_2\,g(t-t_1)g(t'-t_2) \nonumber \\
    &\times& N_{ijkl}(t_1,t_2)\xi^k_0(t_1)\xi^l_0(t_2).
\end{eqnarray}
Note that, while $\expval{W}$ is independent of the noise kernel of gravitons, the second moment $\sigma_W^2$ is not. The noise kernel for an initial thermal distribution of gravitons was computed explicitly in Refs.~\cite{Moreira2023,Cho2022}.

With the expressions for the first and second moments of work, the probability distribution is completely specified. One can now compute
\begin{equation}
    \expval{e^{-\beta W}}=\int dW\,P(W)e^{-\beta W},
\end{equation}
where $\beta=T^{-1}$ is the inverse temperature associated with the gravitons. Since Jarzynski equality holds for the entire system, we must have $\expval{e^{-\beta W}}=e^{-\beta\Delta F}$, where $\Delta F=F(\tau)-F(0)$ is the total free energy difference. $F(t)$ is defined with respect to the Hamiltonian of the system at time $t$, under the assumption that the temperature of the gravitons does not change. 

A direct calculation leads to
\begin{equation}
    \expval{e^{-\beta W}}=e^{-\beta\qty(\expval{W_f}-\beta\sigma_W^2/2)},
\end{equation}
allowing us to identify the free energy difference of the total system as
\begin{equation}
    \Delta F=\expval{W_f}-\frac{\beta}{2}\sigma_W^2,
\end{equation}
which is a sum of a term that depends exclusively on the force protocol $f$ [see Eq. \eqref{First-moment-of-work}] and another that also contains the contribution of gravitons via the noise kernel [see Eqs.~\eqref{Second-moment-of-work} and~\eqref{sigma_ij}].

From this we can directly determine the entropy production $\expval{\Sigma} = \beta\left( \expval{W}-\Delta F\right) = \beta\left( \expval{W_f}-\Delta F\right) $, resulting in
\begin{equation}
\expval{\Sigma} = \frac{\beta^2}{2}\sigma_W^2.
\end{equation}
This result intuitive says that spacetime fluctuations determine the entropy production in the system. There are two interesting limits to discuss here. First, when the gravitational field is turned off, $\sigma_{W}^{2} = 0$, thus implying that there is no entropy production in the system. This is expected since we are dealing with a quantum system that is driven to follow a trajectory in spacetime. Thus, the absence of gravitational waves must imply that there will be no entropy being produced in the system. Secondly, by turning off the driven force, $f=0$, we also have $\expval{\Sigma}=0$. This can be understood because, without the driving force, the system is in equilibrium with the fluctuating environment, and no net work or entropy production should be observed. 

\section{Discussions}
\label{sec:discussion}

Here, we analyze a driven quantum system interacting with a (weak) quantized gravitational field and derive an expression for the entropy production due to gravity by using an integral fluctuation theorem within the framework of decoherent histories. In the regime of low energy scales, perturbations of spacetime are treated quantum mechanically as a quantum field theory, leading to quantum fluctuations of spacetime. These fluctuations are unavoidable, as no system can be shield from gravity ---there is no equivalent of a Faraday cage for gravitational effects. Thus, every system in nature experiences entropy production not only due to processes that potentially drive the system out of equilibrium but also because of the fundamental fluctuations of the spacetime in which the system live. The main result of this work is to characterize these fluctuations. Although decoherence effects were considered before in this context~\cite{Moreira2023}, the thermodynamic entropy was not studied. Thermodynamic considerations are important in two aspects. First, our result helps us to understand how thermodynamics can be formulated in the the case were gravity is present, a problem not yet solved. Secondly, it may point a direction to a quantum statistical description of the gravitational field itself, another important open problem in the field. 

It is important to note that the key aspect of decoherent histories is the need for strong decoherence to eliminate interference among histories. Although crucial for the definition of work, this requirement is not a severe limitation in our case, as decoherence induced by quantum fluctuations of the gravitational field is indeed strong~\cite{Moreira2023,Kanno2021}. Furthermore, quantum particles typically possess additional internal degrees of freedom, which further increases the decoherence rate~\cite{Moreira2023}.

An interesting question for future investigation is the impact of considering a curved spacetime instead of flat spacetime, as was done here. While this introduces many complexities, it should be possible to uncover some fundamental features of entropy production~\cite{Basso2024,Basso2023,Cai2024}.

The result that entropy is produced in quantum systems due to their interaction with gravitational waves could imply decoherence of spatial superpositions of massive particles. This raises the question of how long such superpositions persist. As shown in Ref.~\cite{Moreira2023}, these effects are significant and should be considered when discussing the gravitational field generated by such states~\cite{Belenchia2018,Moller2024}.

Finally, we observed that no entropy is produced when the system is in free fall, in the absence of an external agent. This observation aligns with the equivalence principle. However, this does not imply that quantum spacetime fluctuations are undetected in such a frame, and further investigation is necessary. For example, instead of focusing on the work done by an external agent, one could analyze the work done by the stochastic force resulting from gravitational fluctuations. This distribution of work is not expected to be Gaussian, making the analysis more complex.

\section*{Acknowledgements}
This work was supported by the National Institute for the Science and Technology of Quantum Information (INCT-IQ), Grant No. 465469/2014-0, by the National Council for Scientific and Technological Development (CNPq), Grants No 308065/2022-0, and by Coordination of Superior Level Staff Improvement (CAPES).


\end{document}